# Synthesis of Polyglobalide by Enzymatic Ring Opening Polymerization Using Pressurized Fluids

*Camila Guindani [a*], Wilfred A. G. Jaramillo [b], Graziâni Candiotto [c], Evertan A. Rebelatto [d], Frederico W. Tavares [a,b], José Carlos Pinto [a,b], Papa M. Ndiaye [a,b**], Márcio Nele [a,b***]*

[a] Programa de Engenharia Química / COPPE, Universidade Federal do Rio de Janeiro, Cidade Universitária, CP: 68502, Rio de Janeiro, 21941-972 RJ, Brazil

[b] Programa de Engenharia de Processos Químicos e Bioquímicos / EQ, Universidade Federal do Rio de Janeiro, Cidade Universitária, Rio de Janeiro, 21949-900 RJ, Brazil

[c] Instituto de Química, Universidade Federal do Rio de Janeiro, Cidade Universitária, Rio de Janeiro, 21941-909 RJ, Brazil

[d] Departamento de Engenharia Química e Engenharia de Alimentos, Universidade Federal de Santa Catarina, Florianópolis, 88040-900 SC, Brazil.

[*] Corresponding author: cguindani@peq.coppe.br (C. Guindani)

[**]papa@peq.coppe.ufrj.br (P. M. Ndiaye)

[***]nele@peq.coppe.ufrj.br (M. Nele)

## ABSTRACT

Here, the synthesis of polyglobalide (PGl) by enzymatic ring-opening polymerization (e-ROP) is investigated, using pressurized carbon dioxide ($CO_2$), pressurized $CO_2$ + dichloromethane (DCM), and pressurized propane as solvents. Particularly, the effects of phase equilibrium on the course of e-ROP and PGl final properties are discussed. The partition coefficients of $CO_2$, DCM, propane, globalide and PGl were calculated with help of thermodynamic models, providing proper understanding of monomer partitioning in the reaction system. Reactions performed in pure $CO_2$ resulted in monomer conversions of 100%. Besides, when only one liquid phase was present inside the reactor, PGl samples presented low polydispersities and high average molecular weights. When carried out in $CO_2$ + DCM, e-ROP resulted in lower monomer conversions and PGl samples with higher polydispersities and lower average molecular weights. Finally, reactions carried out in pressurized propane (200 bar) produced PGl samples with the highest average molecular weights among the analyzed products.

## 1. INTRODUCTION

Over the last decades, many efforts have been dedicated to investigation of biodegradable, bioresorbable and biocompatible polymers for applications in biological environments, especially for medical purposes [1,2]. The increasing demand for such materials encourages scientists to design new polymers with tailored properties and develop cleaner manufacturing processes[3–6]. The use of enzymes as biological catalysts to perform ring-opening polymerization reactions (e-ROP) constitutes an excellent example of how the polymerization process can be improved and become cleaner, as demanded by current biomedical applications [7]. Besides, enzymes are normally active under mild temperatures [8], and when immobilized they can be reused [8,9], which have direct impact on the process viability.

In this context, processes that involve the use of pressurized fluids, such as carbon dioxide ($CO_2$), have drawn the attention of researchers to replace the use of organic solvents in e-ROP reactions. Some advantages include the low cost of $CO_2$ and its inherent non-toxic and non-flammable nature [10]. Additionally, $CO_2$ can be readily separated from the final product through system depressurization, making recycling and reuse possible. Compared to other pressurized gases, $CO_2$ is the solvent used most widely to perform enzyme-catalyzed reactions [11]. This is because the temperature range required for its use is compatible with the use of enzymes. Besides, it exhibits transport properties that can accelerate mass transfer in enzymatic reactions [12]. In this case, in order to improve the polymer solubility in the solvent and maintain the low system viscosity (enhancing the rates of mass transfer), co-solvents, such as chloroform and dichloromethane (DCM), can be added to $CO_2$. Because of their low boiling points, these co-solvents can be easily removed from the final material [13]. Nevertheless, $CO_2$ is not the only pressurized gas that can be used as solvent to perform enzymatic reactions. The use of propane has been reported as a promising alternative, exerting a positive effect on the enzymatic activity [11] and presenting a dielectric constant that is similar to the dielectric constant of $CO_2$ [14,15], with the advantage that propane can be used as a pressurized liquid solvent at lower pressures when compared to $CO_2$ [13], which impacts directly in the energy costs of the process.

Polyglobalide (PGl) is a biocompatible and non-toxic polymer that can be produced through polymerization of globalide [16], an unsaturated macrolactone that contains 15 carbon atoms and that has gained increasing attention from the academic community over the last decade. Traditionally, globalide has been used in the perfumery industry as a musk fragrance [17,18]. As the enzymatic ring-opening polymerization of macrolactones (Figure 1) has been investigated over the last decade and shown to be very effective [19–22], manufacture of PGl and its copolymers can constitute an excellent alternative for development of biomedical applications, not only because of their suitable mechanical properties and biocompatibility characteristics [16,23], but also because of the presence of double bonds in the main polymer chain. As a matter of fact, the functionalization of PGl double bonds opens a wide range of possibilities for modification of polymer chains with hydrophilic molecules, proteins, and peptides, which can significantly improve its performance as a biomaterial [6,24–26].

Based on the previous remarks, in the present study the enzymatic synthesis of polyglobalide (PGl) is investigated, using pressurized $CO_2$, a mixture of pressurized $CO_2$ and DCM ($CO_2$+DCM) and pressurized propane as solvents for the first time, and using different globalide to solvent feed ratios. Besides, the effects of phase equilibrium on the course of the e-ROP reactions and final properties of obtained PGl samples (monomer conversion, and average molecular weights, polydispersity, melting temperature and crystallinity of obtained polymer samples) are studied.

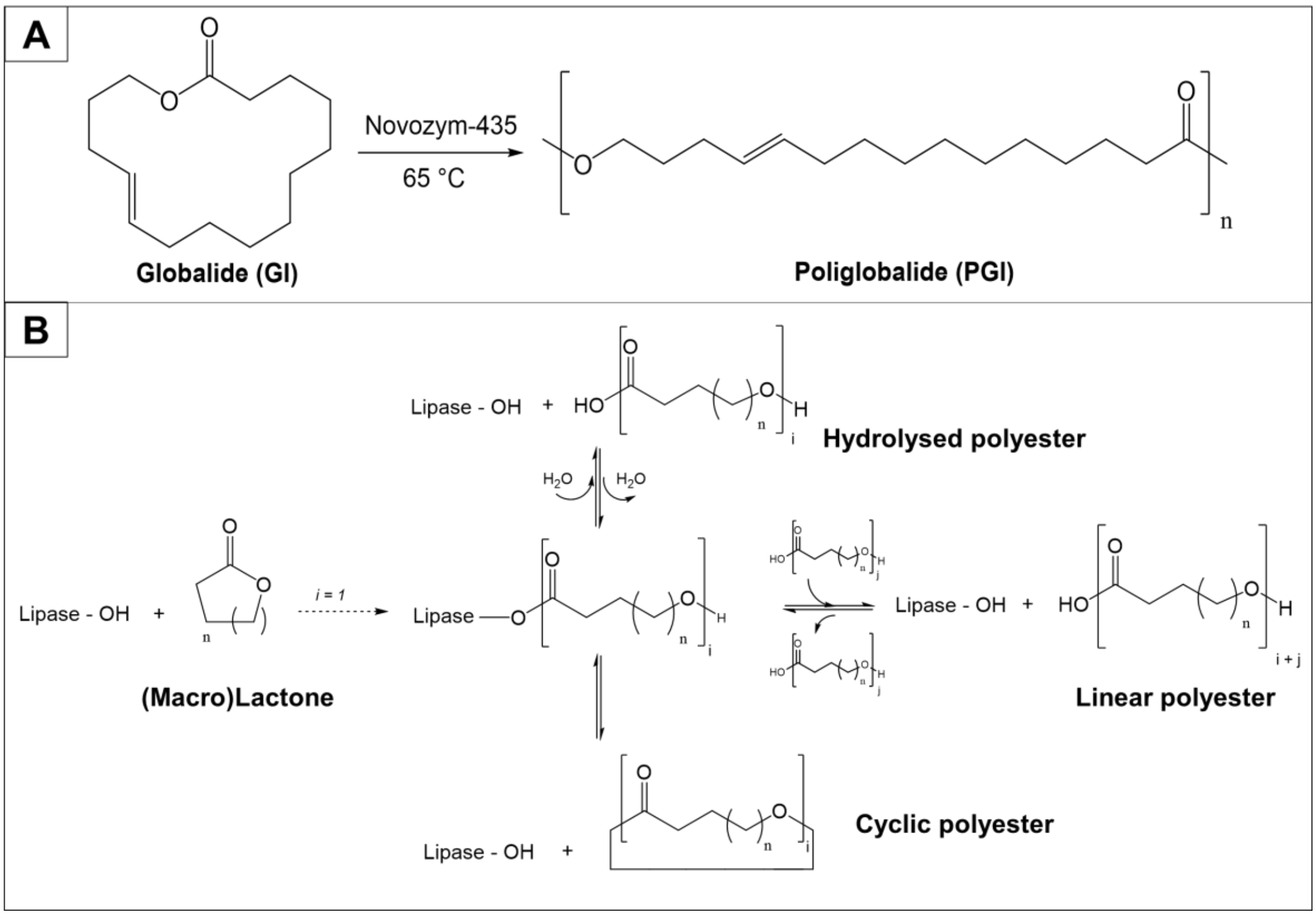

Figure 1. (A) Polymerization of globalide (macrolactone) through enzymatic ring opening reaction; (B) Enzymatic polymerization mechanism for a generic macrolactone monomer.

## 2. MATERIALS AND METHODS

### 2.1. Materials

Dichloromethane P.A. (with purity of 99.8 wt%, DCM) and ethanol P.A. (with purity of 99.8 wt%, EtOH) were purchased from Synth (Brazil). Novozym 435 (commercial lipase B from *Candida antarctica* immobilized on crosslinked polyacrylate beads, with esterification activity of 42 U/g [11]) was kindly donated by Novozymes (Brazil). The monomer globalide (97 wt% purity) was purchased from Symrise (Brazil). Both globalide and the enzymes were dried under vacuum at 70 °C and 60 °C, respectively, and stored in a desiccator over silica and 4 Å molecular sieves [27]. Carbon dioxide (with purity of 99.5 wt%) and propane (with purity of 99.5 wt%) were purchased from White Martins (Brazil) and used as solvents. Unless stated otherwise, chemicals were used as received.

2.2. Enzymatic ring-opening polymerization in pressurized solvents

Polymerization experiments were carried out in a high-pressure variable-volume view cell, equipped with two sapphire windows that allow the visual observation of the reactional medium, an absolute pressure transducer (S model, Swagelok, Ohio, USA) and two syringe pumps (260D and 500D, Teledyne Isco, Lincoln, NE, USA). The cell contains a movable piston that allows the independent control of the reactor pressure. This variable-volume reactor arrangement allows the accurate and independent control of pressure, temperature and composition of the system. The experimental apparatus is based on the high-pressure cell shown in Figure 2. The detailed description of the static–synthetic method adopted here is presented elsewhere, so that the interested reader is encouraged to read these references for additional details [28,29].

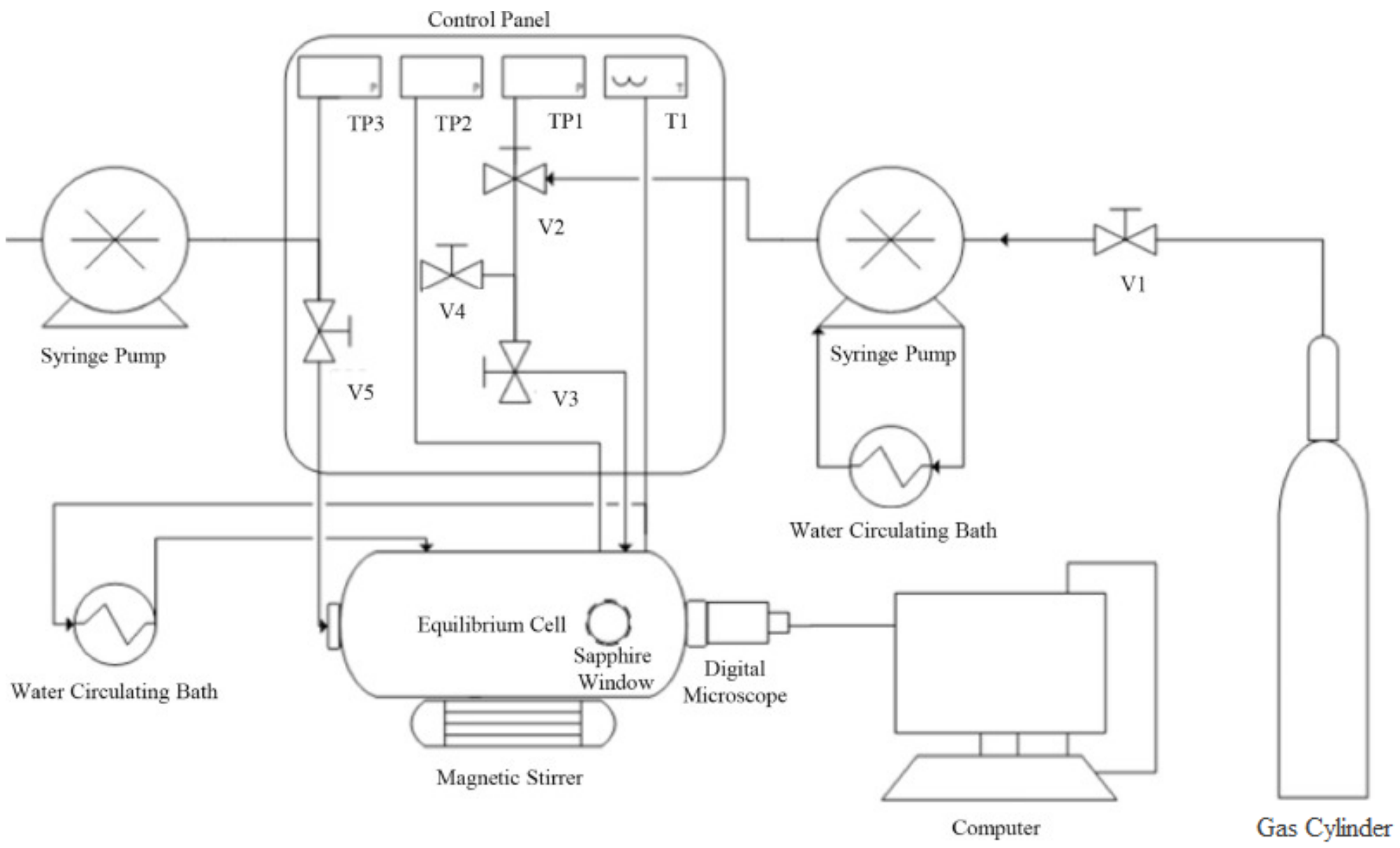

Figure 2. Schematic representation of the high-pressure experimental apparatus (reproduced with permission from Jaramillo et al., 2020 [21]).

Initially, the enzyme (Novozym-435), monomer (globalide) and DCM (co-solvent, when used) were weighed on a precision scale balance (Shimadzu AUX220, Philippines with 0.0001 g accuracy) and introduced into the reactor (equilibrium cell), which was immediately closed. Then, a specified amount of solvent ($CO_2$ or propane) was loaded into the reactor using the syringe pump, with characteristic standard deviation of 0.05 g. The system pressure was then increased until attainment of the specified working pressure. A thermostatic bath was used to supply the heating fluid to the reactor jacket, driving the reactor to the desired reaction temperature. The performances of the reactor temperature and pressure controllers presented characteristic standard deviations of 0.3 ºC and 0.5 bar, respectively. The system was kept under constant stirring with help of a magnetic stirrer and a Teflon$^{TM}$-coated stirring bar. The reaction temperature was kept constant at 65 °C, to attain high enzyme activities [30,31], and the enzyme content was fixed at 5 wt% (in respect to the amount of monomer). Table 1 presents the monomer to solvent feed ratios used in the distinct reaction trials.

Table 1. Reaction conditions in trials performed with globalide (Gl) as monomer, $CO_2$ or propane as solvent, and dichloromethane (DCM) as co-solvent. All reactions were carried out at 65 °C.

| **Run** | **Pressure (bar)** | **$CO_2$:Gl mass ratio** | **Gl:DCM mass ratio** | **Global composition (wt %) $CO_2$/Gl** |
|---|---|---|---|---|
| 1 | 200 | 1:2 | - | 33/67 |
| 2 | | 1:1 | - | 50/50 |
| 3 | | 2:1 | - | 67/33 |
| **Run** | **Pressure (bar)** | **$CO_2$:(Gl+DCM) mass ratio** | **Gl:DCM mass ratio** | **Global composition (wt %) $CO_2$/Gl/DCM** |
| 4 | 200 | 1:2 | 1:2 | 33/22/45 |
| 5 | | 1:1 | 1:2 | 50/17/33 |
| 6 | | 2:1 | 1:2 | 67/11/22 |
| 7 | 200 | 1:2 | 1:1 | 33/33/33 |
| 8 | | 1:1 | 1:1 | 50/25/25 |
| 9 | | 2:1 | 1:1 | 67/17/17 |
| 10 | 200 | 1:2 | 2:1 | 33/44/22 |
| 11 | | 1:1 | 2:1 | 50/33/17 |
| 12 | | 2:1 | 2:1 | 67/22/11 |
| **Run** | **Pressure (bar)** | **Propane:Gl mass ratio** | **Gl:DCM mass ratio** | **Global composition (wt %) Propane/Gl** |
| 13 | 200 | 1:2 | - | 33/67 |
| 14 | 50 | 1:2 | - | 33/67 |

For reactions performed with pressurized $CO_2$ and mixtures of pressurized $CO_2$ and DCM, the pressure was set at 200 bar, based on phase equilibrium data available in the literature [32]. Reactions performed with pressurized propane were carried out with fixed solvent to monomer feed ratio of propane:monomer of 1:2 and pressures of 200 bar and 50 bar, for comparison with results obtained with pressurized $CO_2$. The reaction time was set at 2 hours, usually sufficient to attain high monomer conversions and average molecular weights in enzymatic polymerizations of lactones and macrolactones in pressurized fluids, as described in the literature [21,33,34]. After finishing the polymerization, the obtained raw polymer was solubilized in DCM to allow the removal of the enzymes. The raw polymer solution was then precipitated in cold EtOH. DCM:EtOH was used at the volumetric proportion of 1:6. The obtained polymer suspension was filtrated and dried at room temperature, under vacuum, up to constant weight.

Reaction yield has been traditionally defined as the ratio between the mass of purified polymer and the mass of monomer fed into the reactor. In our case, there is a considerable loss of polymer material inside the reactor due to polymer sticking to the reactor walls, which cannot be recovered efficiently. This way, in order to avoid accounting for such losses in the reaction yield value, the reaction yield was defined as the ratio between the mass of purified polymer and the mass of raw polymer recovered from the reactor, containing non-reacted monomer, oligomers and polymer. Therefore, the reaction yield reflects somehow the efficiency of the purification procedures.

### 2.3. Polyglobalide characterization

*Nuclear Magnetic Resonance - NMR:* $^1$H NMR spectroscopy was performed on a Bruker AC-200F NMR, operating at 400 MHz. Chemical shifts were reported in ppm relative to tetramethylsilane (TMS) (δ=0.00). All samples were solubilized in $CDCl_3$ (δ = 7.27). Polyglobalide samples were analyzed as the raw polymer (without any purification) to determine the monomer conversion. The monomer conversion was determined by comparing the relative intensities of the peaks originating from the methylene protons of the monomer ester and of the polymer ester groups [22,35]. The polyglobalide spectrum shows that the peaks assigned to the methylene protons of globalide (monomer) shifted from 4.14 – 4.07 to 4.07 – 4.03 ppm. The overlapping peaks at 4.07 ppm were separated by applying deconvolution methods using Lorenzian functions [36].

*Gel Permeation Chromatography - GPC:* Number average molecular weight ($M_n$), weight average molecular weight ($M_w$) and polydispersity index (Đ) were determined by gel permeation chromatography. The molecular weight distributions were obtained with a high-performance liquid chromatography equipment (model VE2001, Viscotek, Worcestershire, United Kingdom) equipped with a refractive index detector (model VE3580) and a set of linear columns (Shodex, Tokyo, Japan, models KF-804L and KF-805L, with maximum pore sizes of $1.5 \times 10^3$ and $5 \times 10^3$ Å), using THF as the mobile phase and at room temperature. All samples were solubilized in THF (concentrations of 1.0 $mg.mL^{-1}$) and filtrated through a nylon syringe filter with pores of 0.45 µm. Calibration was carried out with polystyrene standards with average molecular weight ranging from $5 \times 10^2$ to $6.0 \times 10^6$ $g.mol^{-1}$. Analyses were performed at a flow rate of 1.0 $mL.min^{-1}$.

*Differential Scanning Calorimetry - DSC:* Samples of approximately 8 mg of dried polymer were analyzed using a DSC equipment (model Q-1000, TA Instruments, USA) under inert atmosphere (50 $mL{\cdot}min^{-1}$), from 0 to 150 °C at heating rate of 10 $°C.min^{-1}$. The second heating scan was used for determination of thermal transitions in order to ensure consistent thermal history. The first heating scan was performed with heating rate of 20 °C.min$^{-1}$ and cooling rate of 10 °C.min$^{-1}$; therefore, the melting temperature ($T_m$) and the crystallization temperature ($T_{crys}$) of polymer samples were determined during the second heating scan and the second cooling scan, respectively. The degree of crystallinity of the polymer sample was calculated based on its fusion enthalpy ($\Delta H_m$) characterized during the second heating run, and the enthalpy of fusion for a sample of 100% crystalline poly(ε-caprolactone) ($\Delta H_m^0$ = 135.4 J.g$^{-1}$) [37].

## 2.4. Computational methods

The theoretical calculations presented in the present work were based on the Density Functional Theory (DFT). All simulated molecules were generated and analyzed by the software Avogadro [38] version (1.2.0). The calculations associated with the geometrical optimizations of the analyzed molecular structures were performed initially, allowing the determination of the Gibbs free energy ($G$) and the partition coefficient ($P$). The molecular structures of globalide, polyglobalide, $CO_2$, DCM and propane were optimized in the gas-phase, water and n-octanol, and the optimized structures were confirmed as real minima by vibration analysis (no imaginary frequency was detected) [39]. The results were obtained at the reaction conditions (65 °C and 200 bar), using the Becke-three-parameter Lee-Yang-Parr (B3LYP) model as hybrid functional, along with a split-valence double-zeta polarized basis set, based on Gaussian type orbitals (6-31G**) [39]. The Gibbs free energy of solvation in water and n-octanol were computed using the solvation model based on electronic density (SMD) [40]. All DFT calculations were performed using the ORCA 5.0.2 package [41].

## 3. RESULTS AND DISCUSSION

### 3.1. Synthesis of polyglobalide in $CO_2$

The e-ROP of globalide was carried out in pure $CO_2$ and in a mixture of $CO_2$ and DCM (co-solvent). DCM has been frequently used to adjust the polarity of the reaction medium in e-ROP reactions and for production of scaffolds and nanoparticles, with good biocompatibility with different kinds of cells [21,25,42–45]. The absence of toxicity of polymeric devices produced with samples manufactured in presence of DCM is a consequence of its low boiling point (40 °C) [46], which allows the easy separation from the final products [47].

Different monomer to solvent feed ratios were evaluated (Table 1), observing how phase equilibrium changed during the reaction course and how reaction conditions affected the final properties of polyglobalide. All polymerization experiments performed in $CO_2$ were executed under the working pressure of 200 bar and temperature of 65 °C. Remaining parameters were kept constant, as described in the previous section.

Table 2 shows the obtained reaction yields and monomer conversions for each run. As shown in Figure 3, the $^1$H NMR spectra of both globalide and polyglobalide were in good agreement with data available in the literature [16,22,48]. Comparing the monomer conversion values for reactions carried in $CO_2$ (run 1 to 3) and in mixtures of $CO_2$ and DCM (run 4 to 12), one can notice that conversion values were higher in $CO_2$ than in mixtures with the co-solvent. Likewise, yield values followed the same trend.

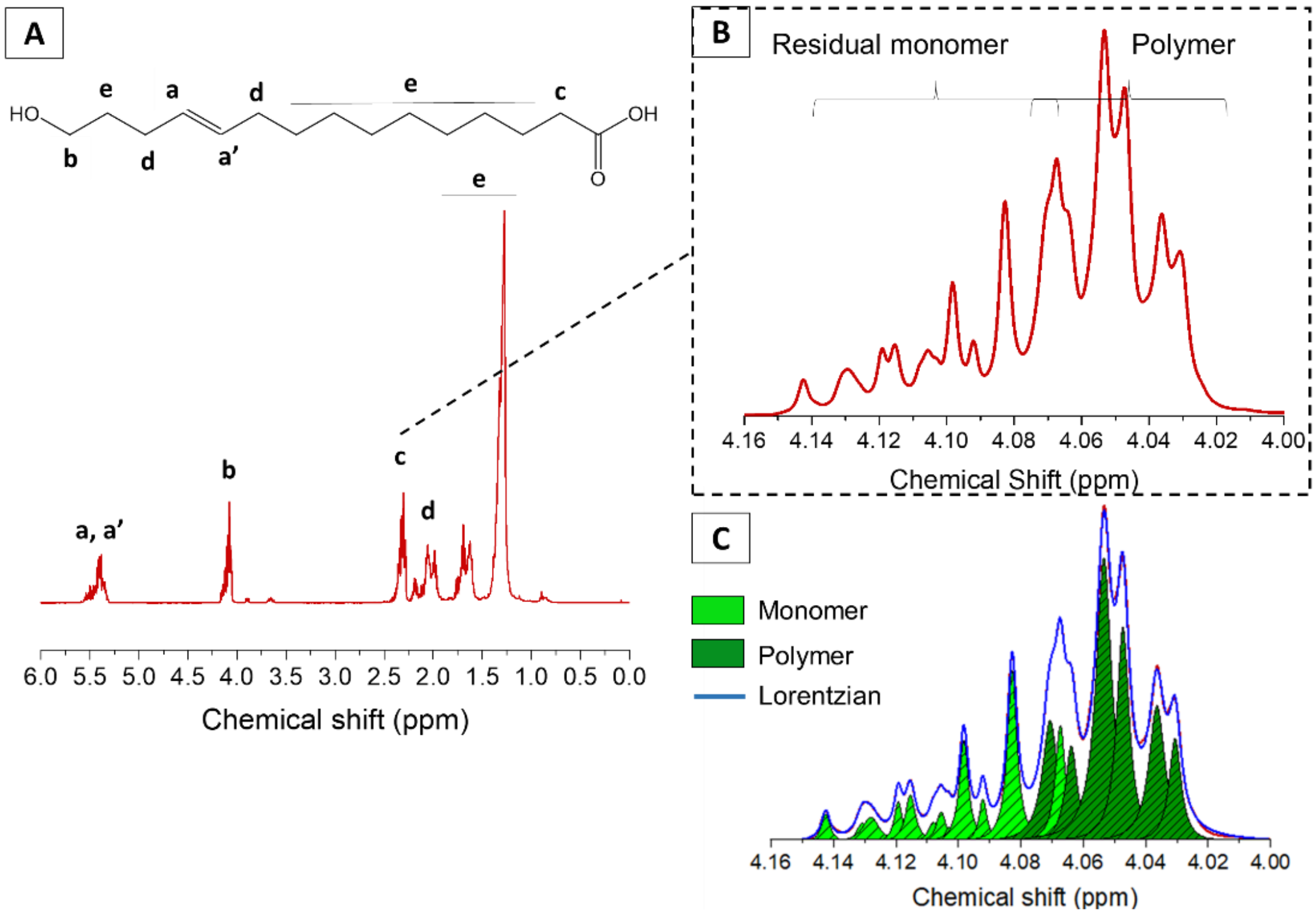

Figure 3. $^1$H NMR spectrum of sample obtained in run 6. (A) PGl structure and peak assignment; (B) expanded view of the methylene peak (4.14 - 4.03 ppm); (C) deconvolution of the overlapping peaks obtained with Lorentzian functions.

Table 2. Monomer conversion and reaction yield values obtained for each run, using $CO_2$ and mixtures of $CO_2$ and DCM as solvents.

| **Run** | **$CO_2$:Gl mass ratio** | **Gl:DCM mass ratio** | **Global composition (wt%) $CO_2$/Gl** | **Conversion** | **Yield** |
|---|---|---|---|---|---|
| 1 | 1:2 | - | 33/67 | 100% | 83% |
| 2 | 1:1 | - | 50/50 | 100% | 85% |
| 3 | 2:1 | - | 67/33 | 100% | 73% |
| **Run** | **$CO_2$:(Gl+DCM) mass ratio** | **Gl:DCM mass ratio** | **Global composition (wt%) $CO_2$/Gl/DCM** | **Conversion** | **Yield** |
| 4 | 1:2 | 1:2 | 33/22/45 | 76% | 61% |
|  | 1:1 | 1:2 | 50/17/33 | 85% | 67% |
| 6 | 2:1 | 1:2 | 67/11/22 | 68% | 70% |
| 7 | 1:2 | 1:1 | 33/33/33 | 79% | 60% |
| 8 | 1:1 | 1:1 | 50/25/25 | 73% | 60% |
| 9 | 2:1 | 1:1 | 67/17/17 | 92% | 68% |
| 10 | 1:2 | 2:1 | 33/44/22 | 72% | 64% |
| 11 | 1:1 | 2:1 | 50/33/17 | 67% | 64% |
| 12 | 2:1 | 2:1 | 67/22/11 | 81% | 73% |

In most cases (except for runs 7, 10, and 11), the addition of DCM to the reaction medium caused the dilution of the monomer, when compared to reactions performed in pure $CO_2$. It is well known [49–51] that the monomer concentration can affect the course of polymerization reactions in at least two important manners: 1) affecting the rates of polymerization by changing the collision frequencies of monomer/oligomers/polymer and the enzyme; 2) and affecting the mass transfer processes, by changing the viscosity of the medium. These effects are somehow competitive, as higher monomer concentrations lead to increase of collision frequencies, but also lead to increase of polymer concentrations and system viscosity, affecting the diffusion of reacting species and, consequently, the reaction rates [50]. Therefore, it could be expected that lower monomer concentrations resulting from increase of DCM concentrations might cause the decrease of monomer conversion and yield values , as reported by Kumar and Gross [52]. However, it was not possible to observe a well-defined trend with the obtained results, being possible to speculate that other effects related to phase equilibrium and monomer partitioning can also affect the course of the reaction.

Figure 4 shows images of the reaction medium at reaction times of approximately 2 h at different reaction conditions. As one can see, for lower $CO_2$ contents (runs 1, 4, 7 and 10), only one liquid phase if formed. It is possible that, due to the smaller amount of solvent ($CO_2$ and $CO_2$ + DCM), in comparison to the other compositions, the melted polymer is solubilizing the solvent, forming an "expanded bulk". As the $CO_2$ increases, a second phase was formed when both pure $CO_2$ and mixtures of $CO_2$ and DCM were used. In the case of pure $CO_2$, the additional phase had the appearance of a vapor phase (rich in $CO_2$), while in the case of the mixtures, the additional phase had the appearance of a second less dense liquid phase, probably rich in $CO_2$ and containing fractions of DCM and monomer [53–55]. In both cases, the polymer material remained in the denser phase, where the formed polymer, oligomers, monomer, enzyme pellets, $CO_2$ and DCM (when used) were present. In the case of reactions performed in $CO_2$, it can be assumed that all monomer remained in the denser liquid phase, given the gaseous aspect of the less dense phase. When DCM was used as co-solvent, the liquid aspect of the lightest phase suggests the possible partitioning of the monomer between the two liquid phases, considering the good affinity between globalide and $CO_2$ [32]. Similar behavior was also reported by Mayer et al., who studied the phase equilibrium behavior of a ternary system containing $CO_2$, ε-caprolactone and dichloromethane [53]. Thus, part of the monomer may possibly remain unreacted in the less dense liquid phase, contributing to the smaller monomer conversion and polymer yield values.

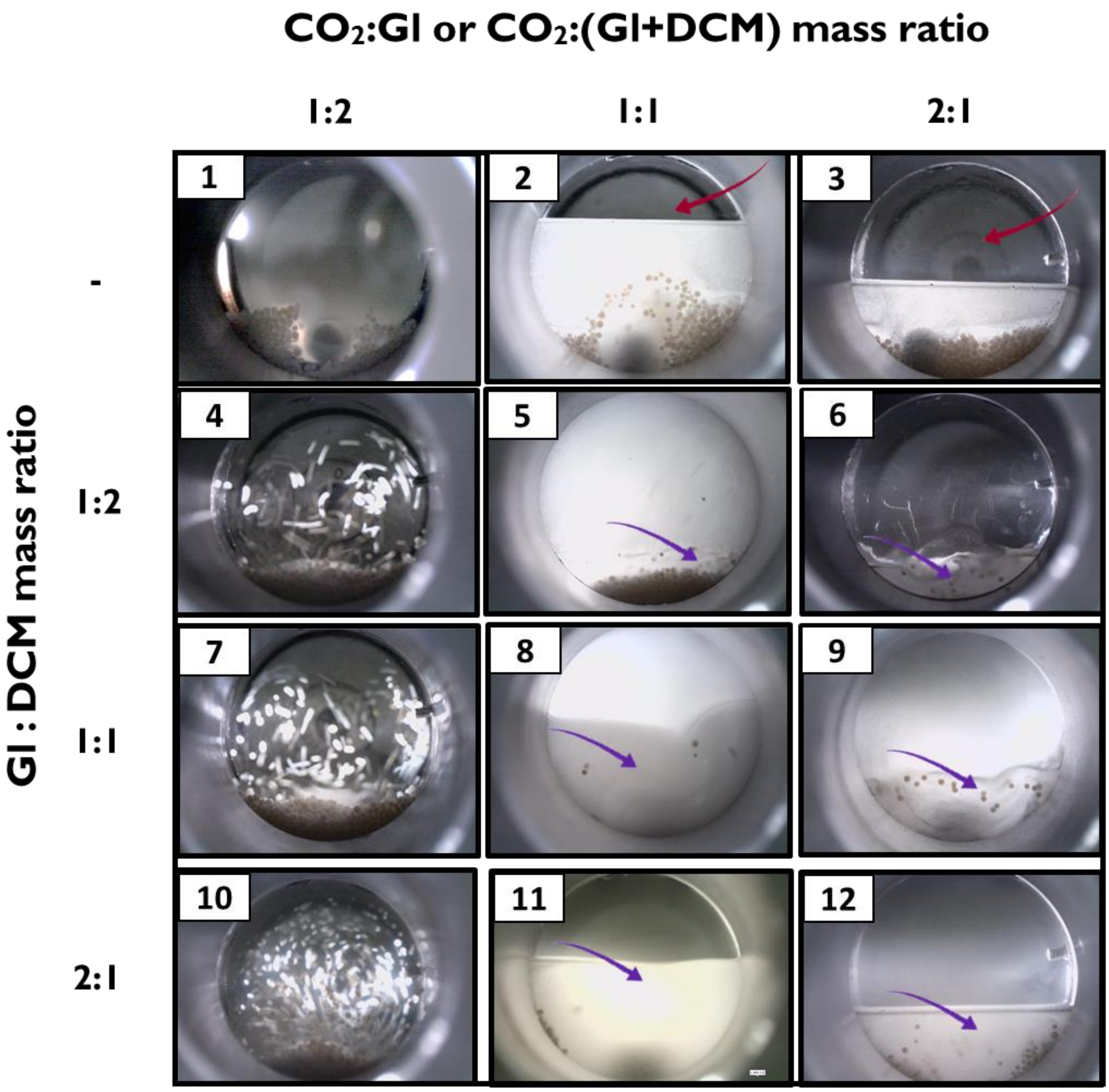

Figure 4. Images obtained during the enzymatic synthesis of PGl using $CO_2$ (1-3) and mixtures of $CO_2$ and DCM (4-12) as solvents. Red arrows indicate the presence of a vapor phase (rich in $CO_2$), while purple arrows indicate the presence of a second liquid phase (denser, rich in PGl). Reactions were carried out at 200 bar and 65 °C during 2 hours.

The previous discussion indicates that different factors may contribute with the reduction of conversion values when DCM is used as co-solvent. For this reason, theoretical prediction of partition coefficients for $CO_2$, propane, DCM, globalide and PGl are reported in the following section and endorse the proposed interpretation of the results.

*3.1.1. Molecular weight distributions*

Average molecular weight and polydispersity values are shown in Table 3. One can observe significant modifications of average molecular weight values of samples synthesized in pure $CO_2$ (runs 1 to 3). As observed in Figure 4, reactions carried out with lower $CO_2$ contents (run 1) did not form a second phase, while obtained polymer samples presented high average molecular weight values and low polydispersities (Đ = 1.66). Compared to run 1, reactions carried out with higher $CO_2$ contents (run 3) formed a second phase (vapor), while the obtained polymer samples presented higher Mw values, lower Mn values and high polydispersities (Đ = 3.19).

The reduction of the $CO_2$ content (and increase of the monomer content) can definitely lead to polymer materials with higher average molecular weights [9,34]. Besides, when the system is kept in a single liquid phase condition (run 1), the monomer concentration decreases uniformly over time and system homogenization is much easier, yielding polymer samples with low values of Đ [50]. However, when a second vapor phase is formed (runs 2 and 3), the monomer concentration in the reacting phase is also expected to increase, leading to more complex variation of monomer concentration and reaction rates over time. Additionally, reduction of the volume of the reacting phase can cause the increase of viscosity and more difficult homogenization of the reacting system, especially considering that the agitation promoted by the stirring bar is easily impaired. This fact can explain the higher Đ values [56]. Figure 5 shows the molecular weight distributions of obtained samples for comparative purposes.

Table 3. Mn, Mw and Đ values for PGl samples synthesized in $CO_2$ and mixtures of $CO_2$ and DCM.

| Run | $CO_2$:Gl mass ratio | Gl:DCM mass ratio | Global composition (wt%) $CO_2$/Gl | Mn (Da) | Mw (Da) | Đ |
|---|---|---|---|---|---|---|
| 1 | 1:2 | - | 33/67 | 15,222 | 25,271 | 1.66 |
| 2 | 1:1 | - | 50/50 | 9,819 | 31,367 | 3.19 |
| 3 | 2:1 | - | 67/33 | 8,362 | 30,659 | 3.67 |
| **Run** | **$CO_2$:(Gl+DCM) mass ratio** | **Gl:DCM mass ratio** | **Global composition (wt%) $CO_2$/Gl/DCM** | **Mn (Da)** | **Mw (Da)** | **Đ** |
| 4 | 1:2 | 1:2 | 33/22/45 | 6,082 | 14,721 | 2.42 |
| 5 | 1:1 | 1:2 | 50/17/33 | 8,580 | 21,455 | 2.50 |
| 6 | 2:1 | 1:2 | 67/11/22 | 8,300 | 18,908 | 2.28 |
| 7 | 1:2 | 1:1 | 33/33/33 | 6,315 | 13,071 | 2.07 |
| 8 | 1:1 | 1:1 | 50/25/25 | 8,381 | 21,691 | 2.59 |
| 9 | 2:1 | 1:1 | 67/17/17 | 9,560 | 22,662 | 2.37 |
| 10 | 1:2 | 2:1 | 33/44/22 | 10,323 | 30,076 | 2.91 |
| 11 | 1:1 | 2:1 | 50/33/17 | 12,704 | 29,043 | 2.29 |
| 12 | 2:1 | 2:1 | 67/22/11 | 16,056 | 36,345 | 2.26 |

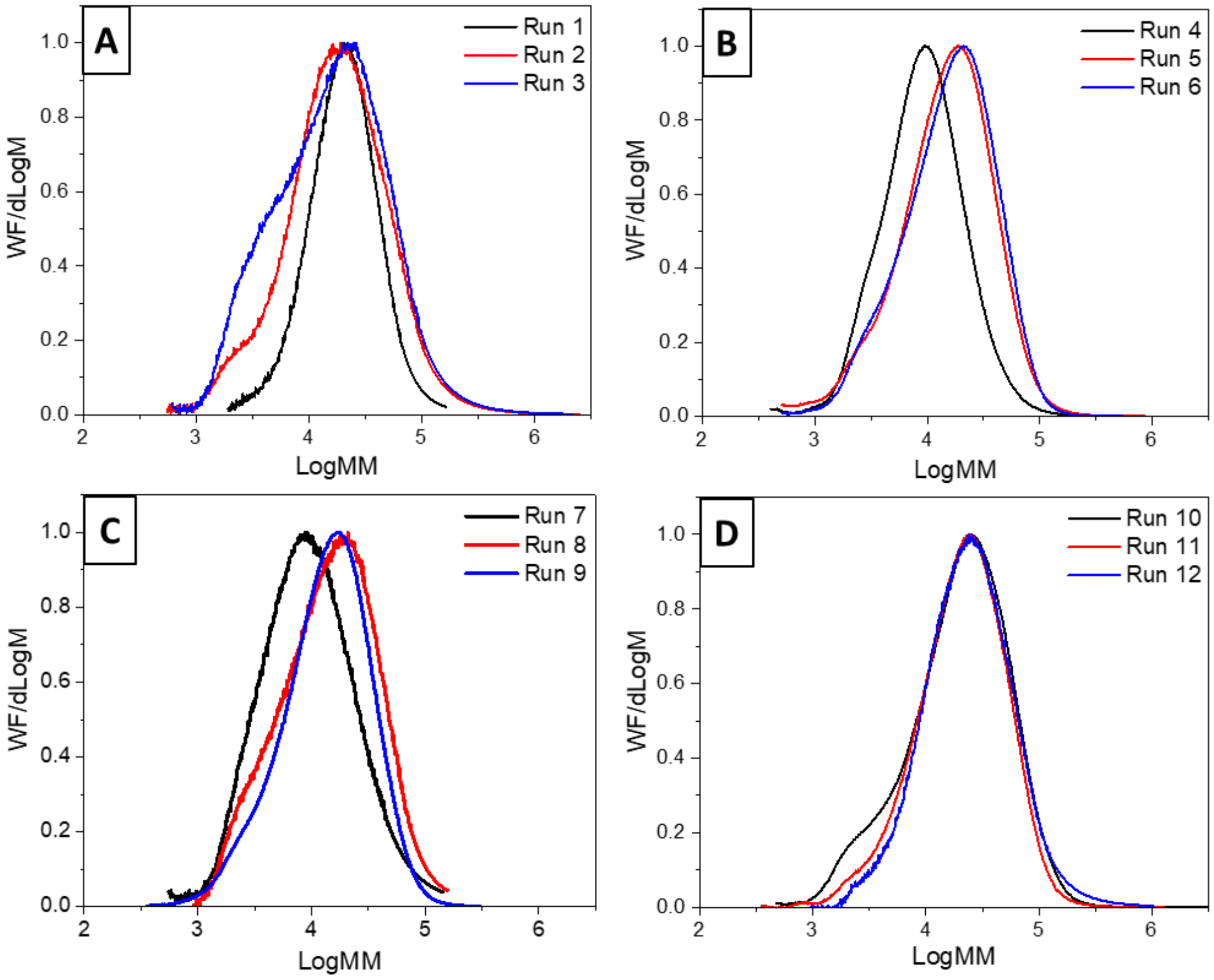

Figure 5. Molecular weight distributions of PGl samples obtained in (A) pure $CO_2$ and mixtures of $CO_2$ and DCM with monomer to co-solvent feed ratios of (B) monomer:DCM = 1:1, (C) monomer:DCM = 1:2 and (D) monomer:DCM = 2:1.

When DCM was used as a co-solvent, the volume of the second phase also increased with the $CO_2$ content, although the Mn and Mw values increased, while Đ values remained almost unchanged. This probably happened because the presence of DCM in the denser phase reduced the viscosity of the system, allowing the better homogenization of the system throughout the reaction time, leading to lower polydispersity index values in comparison to run 3 [49,56].

Finally, when comparing different co-solvent to monomer feed ratios for fixed amounts of $CO_2$, it can be observed that experiments performed with smaller amounts of DCM led to higher values of Mn and Mw, as it might already be expected when the monomer concentration controls the rates of chain growth. Particularly, this can be observed for run 12, which provided a Mn value that was similar to the one obtained for run 1 and an even higher Mw value. Therefore, it can be conjectured that the amount of DCM at this condition was sufficient to maintain the homogenization of the reaction medium without causing significant reduction of the polymerization rate.

#### 3.1.2. *Thermal analyses*

First, we have performed a Thermogravimetric (TGA) analysis in order to access the weight loss curve caused by thermal decomposition of PGl (see Support Information, Figure S1). Then, the thermal behaviors of the obtained PGl samples were evaluated through DSC analyses, as shown in Table 4. It can be observed that the samples presented degrees of crystallinity ranging from 50% to 70%, and $\Delta H_m$ values that were consistent with data reported in the literature for polyesters [21,57,58]. The variations observed in the degree of crystallinity observed for the evaluated conditions are probably a result of a thermal lag effect [59], being not related to the synthesis conditions. At two reaction conditions, obtained polymer samples presented double melting peaks (run 4, Figure 6B and run 7, Figure 6C). The occurrence of the so-called “multiple fusion behavior” is very often observed for semi-crystalline polymers [60]. This behavior can be related to the presence of small and imperfect crystals that can present sufficient mobility to rearrange themselves into more stable and larger crystals with the increasing temperature during the analysis, which is known as a “fusion-recrystallization” mechanism [61]. It can be noted that double melting peaks were detected for samples that presented the lowest average molecular weights (Table 2 and Figure 4). Samples with low average molecular weight frequently show double melting peaks because they usually form less perfect crystal structures, due to the presence of higher concentrations of end groups, which can lead to increase of the free volume and can prevent the appropriate packing of the chains, causing deformation of the lamellar thickening process [62].

Table 4. Thermal properties of PGl samples synthesized at different reaction conditions, using $CO_2$ as solvent.

| **Run** | **$\Delta H_m$ (J/g polymer)** | **$X_{crys}$ (%)** | **$T_{crys}$ (°C)** | **$T_{m1}$ (°C)** | **$T_{m2}$ (°C)** |
|---|---|---|---|---|---|
| 1 | 73.09 | 54% | 32 | 46 | - |
| 2 | 74.17 | 55% | 34 | 47 | - |
| 3 | 74.21 | 55% | 36 | 49 | - |
| 4 | 90.85 | 67% | 38 | 47 | 43 |
| 5 | 81.41 | 60% | 34 | 48 | - |
| 6 | 76.65 | 57% | 35 | 48 | - |
| 7 | 82.3 | 61% | 36 | 46 | 42 |
| 8 | 81.01 | 60% | 34 | 47 | - |
| 9 | 78.04 | 58% | 35 | 48 | - |
| 10 | 71.95 | 53% | 32 | 46 | - |
| 11 | 74.01 | 55% | 33 | 46 | - |
| 12 | 68.61 | 51% | 32 | 46 | - |

$T_{crys}$: Crystallization temperature; $T_{m1}$: Temperature of the first melting peak; $T_{m2}$: Second melting peak temperature; $\Delta H_m$: Enthalpy of fusion; $X_{crys}$: Degree of crystallinity, calculated from the enthalpy of fusion for a sample of 100% crystalline poly(ε-caprolactone) ($\Delta H_m^0$ = 135.4 J.g$^{-1}$) [37].

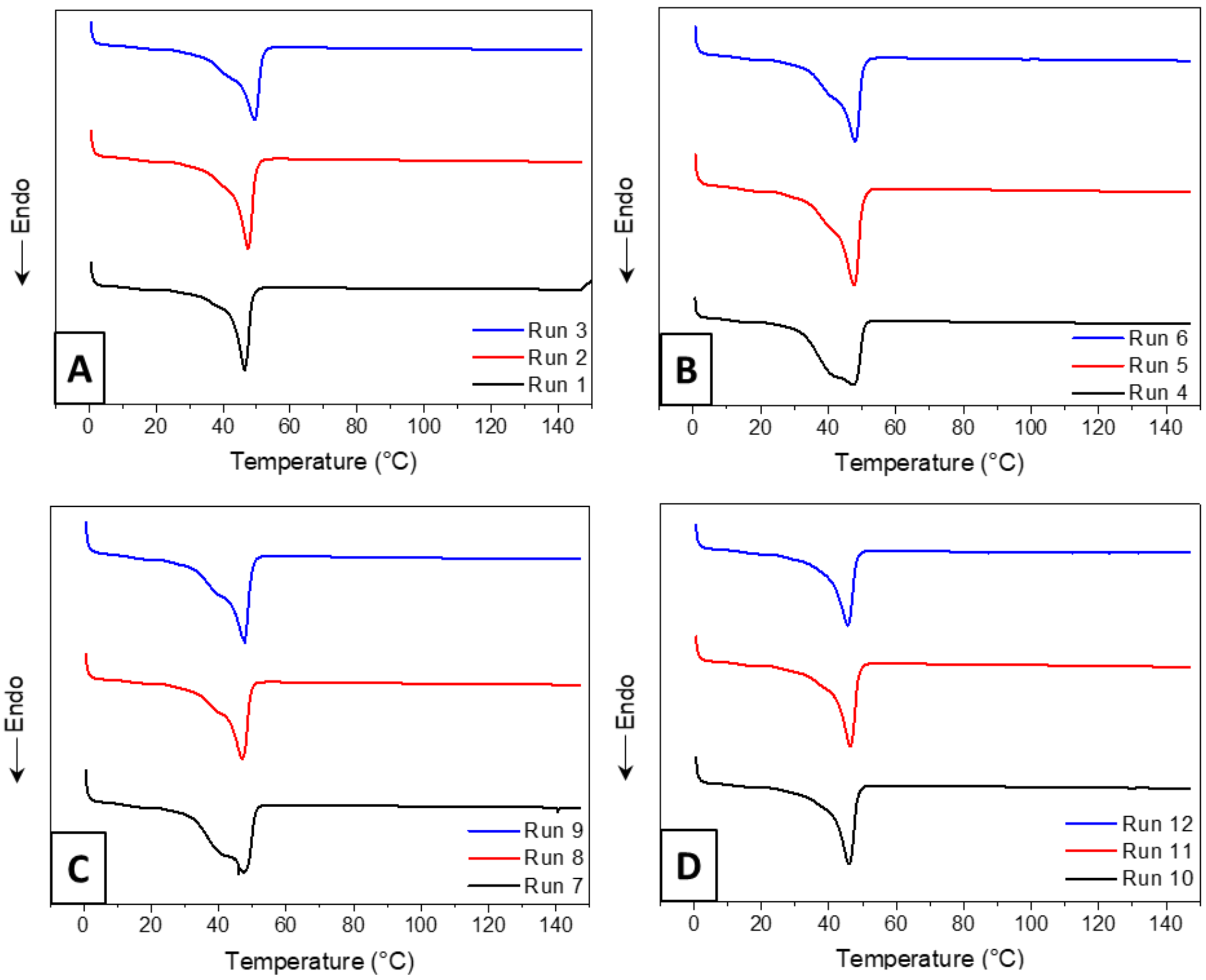

Figure 6. DSC thermograms of PGl samples synthesized in (A) pure $CO_2$ and mixtures of $CO_2$ and DCM with monomer to co-solvent feed ratios of (B) monomer:DCM = 1:1, (C) monomer:DCM = 1:2 and (D) monomer:DCM = 2:1.

## 3.2. <u>Synthesis of polyglobalide in pressurized propane</u>

Propane was used as solvent because of its positive effect on enzymatic activity [11] and similar dielectric constant (polarity), when compared to $CO_2$ (~1.6) [14,15]. Besides, propane has lower compressibility than $CO_2$, which means that propane can be used at lower operation pressures than $CO_2$, which can lead to lower processing costs. For example, propane exhibits density of 0.436 g / mL at 65 °C and 50 bar, while $CO_2$ attains similar density values at 128 bar [13].

Polymerization experiments in pressurized propane were performed under working pressures of 200 bar (for comparison with the experiments carried out in $CO_2$) and 50 bar. All other parameters, such as temperature, enzyme content and reaction time, were kept constant, as described in the previous section. The solvent to monomer feed ratio was kept constant and equal to 1:2, once this was the condition that provided higher conversion and yield values in $CO_2$ and allowed the manufacture of polymer samples with narrow molecular weight distribution (low Đ) and high Mn and Mw values.

Table 5 shows the monomer conversion and reaction yield values obtained in the experiments performed with pressurized propane as solvent. For both runs, the values obtained for monomer conversion and yield were lower than for $CO_2$. One of the possible reasons for this behavior is the higher viscosity of pressurized propane (69.180 μPa.s at 50 bar / 65 °C and 95.428 μPa.s at 200 bar / 65 °C) in comparison to $CO_2$ (56.101 μPa.s at 200bar / 65 °C) [13]. Besides, it is important to mention that pressurized $CO_2$ is reported to act as a plasticizer for polymers, reducing its viscosity [63,64]. As mentioned in the previous sections, it was possible to observe through the sapphire window that the high viscosity of the system can affect the mass transfer processes and reduce the collision frequencies between monomer and enzyme molecules, leading to smaller monomer conversion and reaction yield values. For instance, when comparing reaction trials carried out at 50 bar (run 14) and 200 bar (run 13) (Figure 7), it can be observed that the lower viscosity of the solvent can cause the slight increase of monomer conversion values, although the opposite behavior was observed for yields, which can also be related to the polymer purification process, as shown in the following section.

Table 5. Monomer conversion and reaction yield values obtained for runs performed in pressurized propane.

| Run | Pressure (bar) | Propane:Gl mass ratio | Global composition (wt%) Propane/Gl | Conversion | Yield |
|---|---|---|---|---|---|
| 13 | 200 | 1:2 | 33/67 | 81% | 72% |
| 14 | 50 | 1:2 | 33/67 | 89% | 65% |

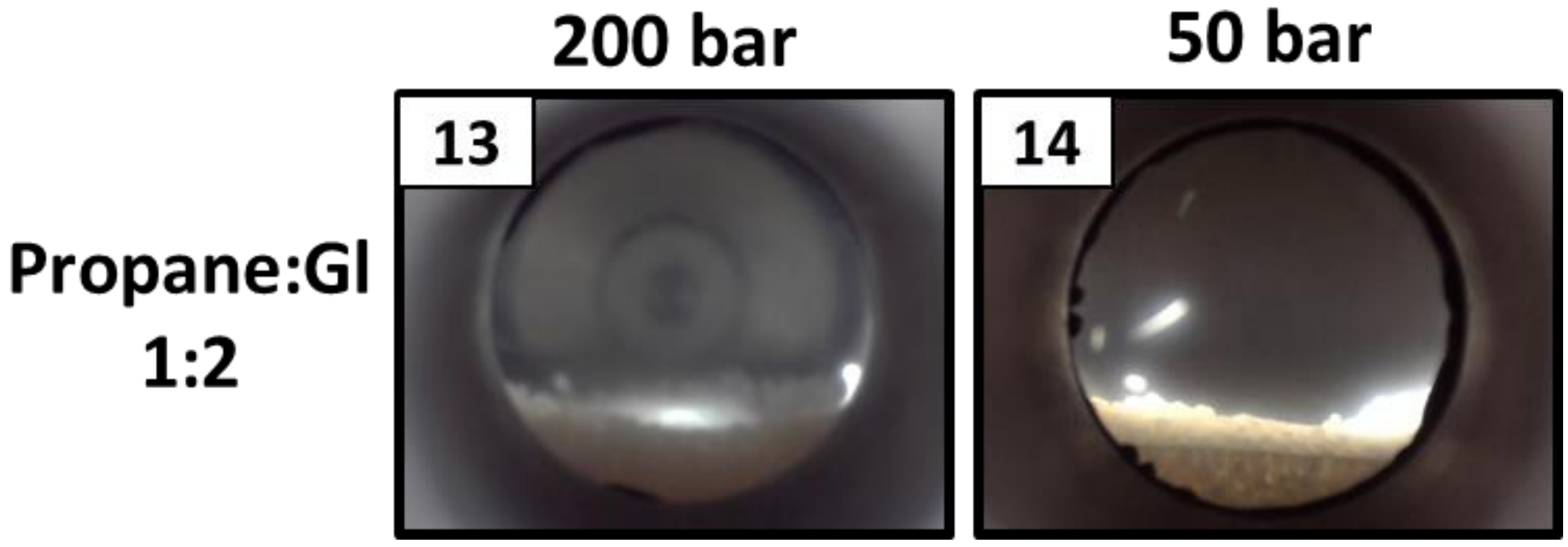

Figure 7: Images obtained during the enzymatic synthesis of PGl using pressurized propane at 200 bar (13) and 50 bar (14).

3.2.1. *Molecular weight distributions*

Average molecular weights and polydispersities of polymer samples prepared in runs 13 and 14 are presented in Table 6, while the respective molecular weight distributions are presented in Figure 8. Compared to run 1 carried in $CO_2$ at similar reaction conditions, the sample prepared in propane (run 13) presented slightly higher Mn value (15,222 Da in $CO_2$ and 16,599 Da in propane) and much higher Mw value (25,271 Da in $CO_2$ and 39,995 Da in propane), which can be advantageous for many applications [62,65,66] and reveal that the polymer chains are longer when synthesized in pressurized propane. Similar results were reported by Rosso et al. [34], when poly(ε-caprolactone) samples prepared via e-ROP in $CO_2$ and liquefied petroleum gas (LPG) were compared to each other. This can be due to the positive effect of propane over the enzyme activity, as reported in the literature [11]. However, this can also be due to mass transfer effects related to the different viscosities of the reaction media and to the good affinity between globalide and propane, which leads to good solubilization of the formed oligomers and polymer species, as shown in the following section.

Finally, when carried out at 50 bar, the polymerization reaction performed in propane yielded a polymer with higher polydispersity index (Đ = 3.31) and lower Mn and Mw values, when compared to the values reported for 200 bar. The molecular weight distributions of these polymer samples revealed that run 14 provided samples with a bimodal distribution that was shifted towards lower molecular weight values, when compared to run 13, indicating the formation of shorter chains and the existence of heterogeneous reaction conditions. The decrease of the propane density, and consequent decrease of the solvation power (0.436 g/mL at 50 bar / 65 °C and 0.488 g/mL at 200 bar / 65 °C) might be one of the possible explanations, allied to an insufficient degree of agitation promoted by the stirring bar, although more fundamental investigations are necessary to present an unequivocal explanation for this behavior.

Table 6. Mn, Mw and Đ values for PGl samples synthesized in pressurized propane.

| Run | Pressure (bar) | Propane:Gl mass ratio | Global composition (wt%) Propane/Gl | Mn (Da) | Mw (Da) | Đ |
|---|---|---|---|---|---|---|
| 13 | 200 | 1:2 | 33/67 | 16,599 | 39,995 | 2.41 |
| 14 | 50 | 1:2 | 33/67 | 7,747 | 25,606 | 3.31 |

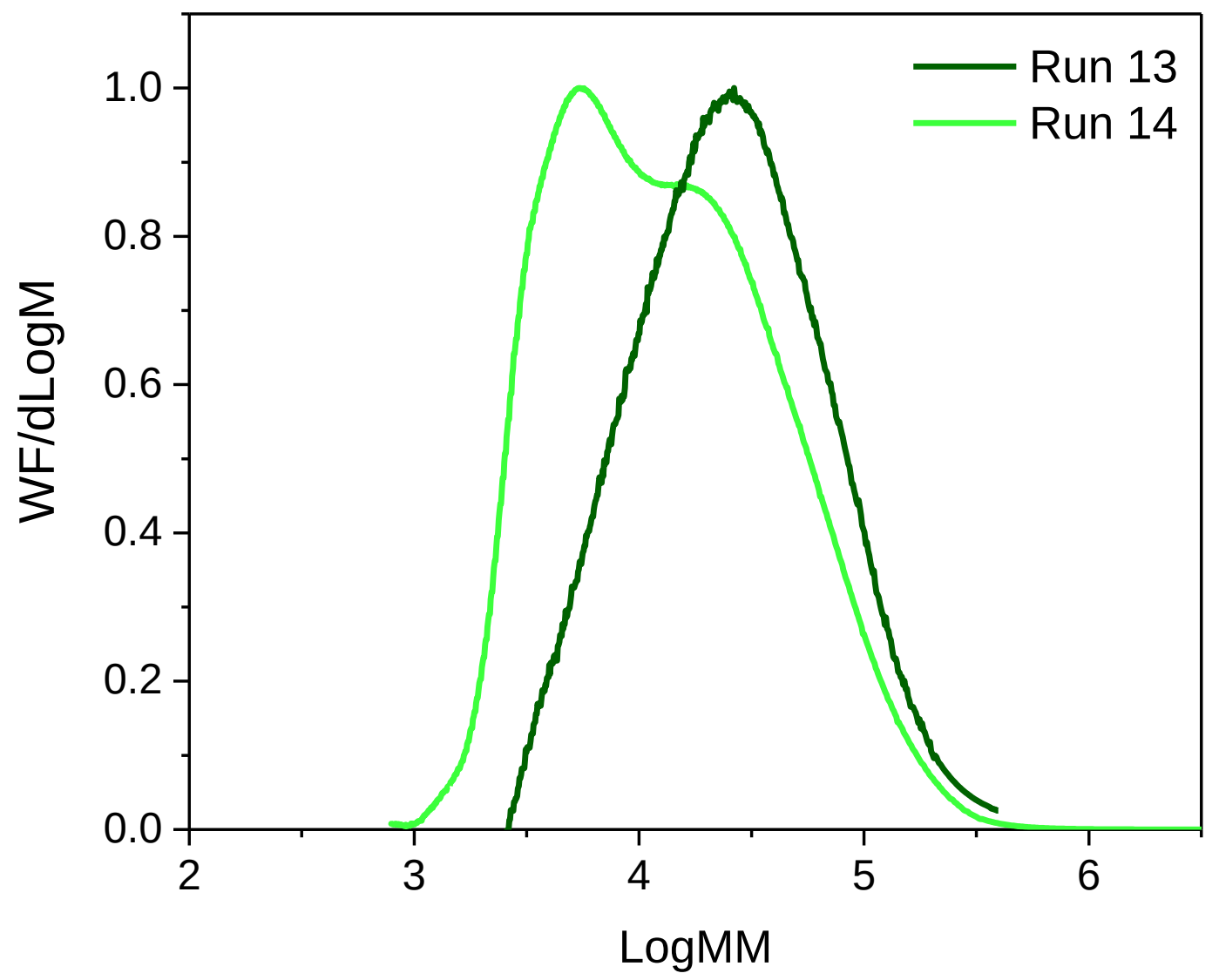

Figure 8. Molecular weight distributions of PGl samples obtained in pressurized propane at 200 bar (run 13) and 50 bar (run 14).

### 3.2.2. *Thermal analyses*

Table 7 presents the thermal properties of PGl samples prepared in pressurized propane at 200 bar and 50 bar, while Figure 9 presents the DSC thermograms of the samples. The degrees of crystallinity obtained for polymer samples prepared in runs 13 and 14 were very similar to the ones obtained in pure $CO_2$. The same can be said for $T_m$, $T_{crys}$, and $\Delta H_m$.

All values presented here are consistent with previously reported values published in the literature for polyesters [21,57,58].

Table 7. Thermal properties of PGl samples synthesized in pressurized propane at 200 bar (13) and 50 bar (14).

| **Run** | **$\Delta H_m$ (J/g polymer)** | **$X_{crys}$ (%)** | **$T_{crys}$ (°C)** | **$T_{m1}$ (°C)** | **$T_{m2}$ (°C)** |
|---|---|---|---|---|---|
| 13 | 68.14 | 50 | 32 | 47 | - |
| 14 | 72.78 | 54 | 35 | 49 | - |

$T_{crys}$: Crystallization temperature; $T_{m1}$: Temperature of the first melting peak; $T_{m2}$: Second melting peak temperature; $\Delta H_m$: Enthalpy of fusion; $X_{crys}$: Degree of crystallinity, calculated from the enthalpy of fusion for a sample of 100% crystalline poly(ε-caprolactone) ($\Delta H_m^0$ = 135.4 J.g$^{-1}$) [37].

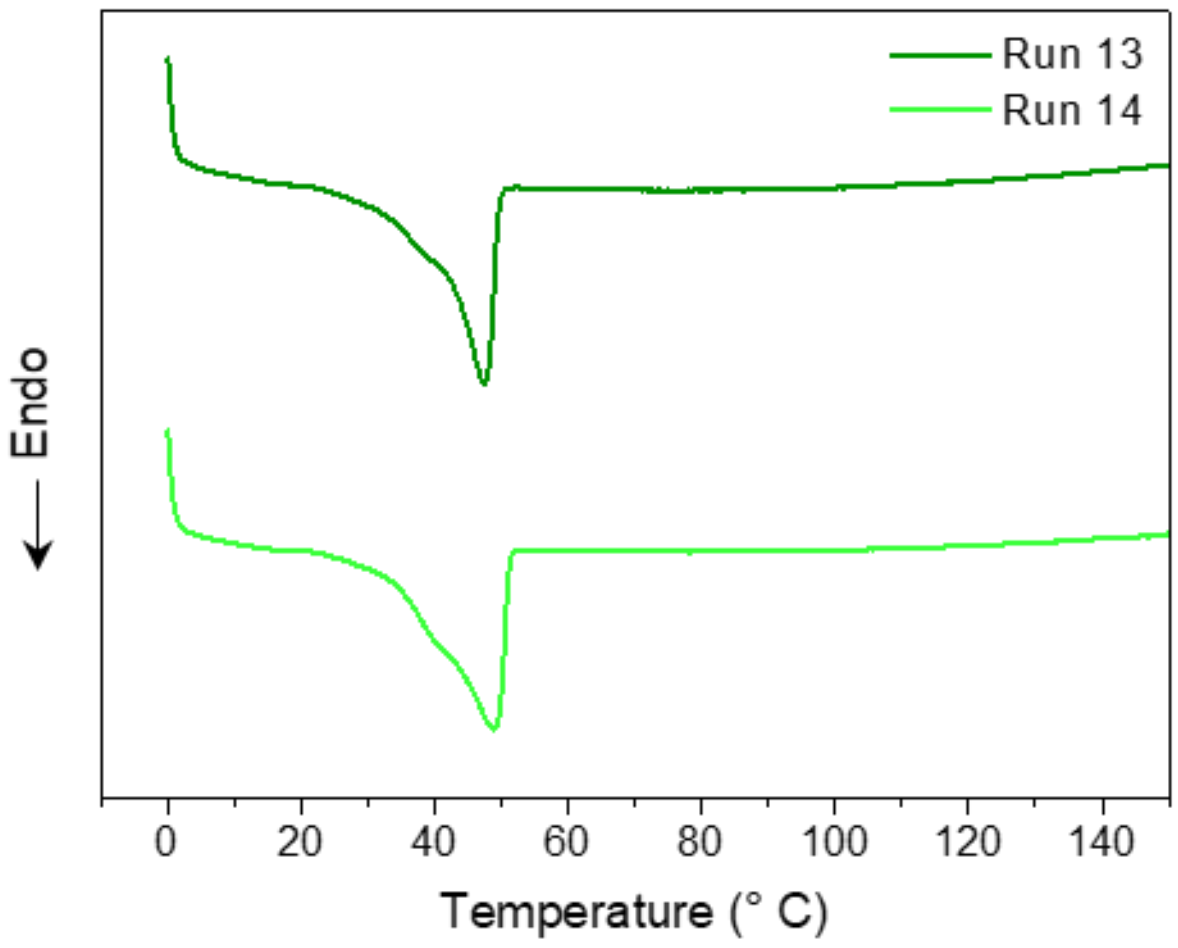

Figure 9. DSC thermograms of PGl samples prepared in pressurized propane at 200 bar (run 13) and 50 bar (run 14).

### 3.3. Theoretical considerations

DFT calculations were performed to estimate some thermodynamic properties of mixtures containing globalide, polyglobalide, and the solvents $CO_2$, DCM and propane, such as the Gibbs free energy of solvation and the partition coefficients of each individual component. These properties can provide better understanding about the interactions among the system components and how thermodynamic partitioning can affect the reaction course.

The partition coefficient ($P$) can be defined as the ratio between the concentration of a solute in two phases of a mixture that contains two immiscible solvents at equilibrium [67]. It must be noted that partition coefficients depend on the Gibbs free energy of solvation. The theoretical prediction of partition coefficients for n-octanol/water mixtures ($log\ P^{O/W}$) were calculated according to Equation (1) [5,68], where the Gibbs free energy ($G$) and the Gibbs free energy of solvation ($\Delta G_{solv}$) were estimated with help of the DFT calculations. The superscripts $W$ and $O$ represent respectively the water and n-octanol solvents, while $R$ is the ideal gas constant (8.314 $J/K^{-}mol^{-1}$) and $T$ is the reaction temperature (338.15 K).

$$\log P^{O/W} = \frac{\Delta G^{W}_{solv} - \Delta G^{O}_{solv}}{2.303RT} \tag{1}$$

The calculated $\log P^{O/W}$ values are presented in Table 8, while the Gibbs free energy ($G$) and the Gibbs free energy of solvation ($\Delta G_{solv}$) are presented in Table S1 of the support information. The partition coefficient values of the polymer chains were calculated based on an approximation that considered only 2 repeating units.

Table 8. $Log\ P^{O/W}$ of $CO_2$, DCM, propane, globalide and PGl calculated at 65 °C and 200 bar, using DFT/B3LYP/6-31G** with water and n-octanol solvents in SMD model.

| Compound | $\log P^{O/W}$ | |
|---|---|---|
| | Calculated | Literature |
| $CO_2$ | 1.97 | 2.00 [69] |
| DCM | 1.42 | 1.25 [46] |
| Propane | 2.15 | 2.36 [70,71] |
| Globalide | 5.69 | 5.70 [72] |
| PGl (1 un.) | 6.39 | - |
| PGl (2 un.) | 6.82 | - |

The calculated $log\ P^{O/W}$ for $CO_2$, DCM, propane and globalide are consistent with data reported previously in the literature, indicating that the calculation procedure was implemented adequately. The monomer globalide presented $\log P^{O/W}$ value of 5.69, revealing the strong hydrophobic character. After the ring-opening reaction, PGl (1 un.) assumes a $\log P^{O/W}$ of 6.39 and its value increase as the number of repeat units increase, as one can observe for PGl (2 un.), where $\log P^{O/W}$= 6.82. Therefore, the hydrophobic character of the product increases, as the number of repeat units increases. Due to the high computing costs involved, it is not practicable to calculate $\log P^{O/W}$ of PGl containing more than 2 repeat units. However, it is well known that hydrophobic molecules, including polyesters [73,74], can reach $\log P^{O/W}$ > 10 [75], so it is reasonable to assume that PGl can reach similar values. Considering that globalide has much lower molecular weight (and, therefore, higher mobility) in comparison to PGl, and an intermediate $\log P^{O/W}$ value (compatible with $CO_2$, DCM and PGl), it is probable that globalide is partitioned between the two phases of the systems where liquid-liquid equilibrium was observed, as assumed in Section 3.1, explaining why these systems presented conversion values lower than 100% .

The calculated $\log P^{O/W}$ values also endorses the good performance of pressurized propane as solvent, producing high molecular weight PGl. Propane presented higher $\log P^{O/W}$ value when compared to $CO_2$ and DCM, indicating its greater affinity to globalide and PGl, which implies in higher solubility. These results are consistent with the results reported by Kumar and Gross [52], who obtained poly(ε-caprolactone) (PCL) samples of higher average molecular weights by e-ROP when using solvents with $\log P^{O/W}$ between 1.9 and 4.5, suggesting that this is a consequence of the higher solubility of PCL in solvents with lower polarity, allied to its capacity to preserve the conformation of the enzymes, keeping its activity.

## 4. CONCLUSIONS

For the first time, results regarding the enzymatic synthesis of polyglobalide (PGl) in $CO_2$, mixtures of $CO_2$ and DCM (dichloromethane) and pressurized propane are reported in the literature. Different monomer to solvent feed ratios were investigated, and the effects of reaction conditions on monomer conversion, reaction yield, average molecular weights, and thermal properties were evaluated. Reactions carried out in $CO_2$ resulted in monomer conversion of 100% and yielded PGl samples with low polydispersity index (Đ) and high average molecular weights (Mn = 15,222 Da, Mw = 25,271 Da, Đ = 1.66). When DCM was used as co-solvent, lower monomer conversion values were obtained, while polymer samples with higher polydispersities, Mn and Mw values were produces (Mn =16,056 Da, Mw = 36,345 Da). Moreover, reactions performed in pressurized propane at 200 bar provided polymer samples with the highest Mn, Mw and polydispersity values (Mn = 16,599 Da, Mw = 39,995 Da, Đ = 2.41). As discussed throughout the text, the obtained results can be affected simultaneously by a large number of factors, including the variation of monomer concentration, the modification of the viscosity of the reacting system (and, therefore, of mass transfer rates) and the occurrence of phase separation when the $CO_2$ content is increased. The calculation of partition coefficients reinforced the assumption that monomer partitioning between the different liquid phases during polymerization can significantly affect the course of the reaction and can explain the better performance of the propane solvent during the PGl synthesis. Finally, thermal analyses indicated that the melting temperature, the enthalpy of fusion, and the degree of crystallinity of obtained PGl samples were similar to values reported for other polyesters. Besides, polymer samples with low average molecular weights presented typical double melting point behavior, caused by the increase of the number of terminal groups and, consequently, of the free volume of the system.

## ACKNOWLEDGMENTS

The authors thank the Instituto de Macromoléculas Eloisa Mano (IMA-UFRJ) for the DSC analyses, and LABRMN (IQ-UFSC) for the NMR analyses. C. Guindani thanks FAPERJ (Fundação Carlos Chagas Filho de Amparo à Pesquisa do Estado do Rio de Janeiro), process number E-26/201.911/2020 and E-26/201.912/2020, for the financial support. G. Candiotto gratefully acknowledges the computational support of Núcleo Avançado de Computação de Alto Desempenho (NACAD/COPPE/UFRJ) and Sistema Nacional de Processamento de Alto Desempenho (SINAPAD), and FAPERJ (Processo E-26/200.008/2020) for financial support. The authors also thank CAPES (Coordenação de Aperfeiçoamento de Pessoal de Nível Superior) and CNPq (Conselho Nacional de Desenvolvimento Científico e Tecnológico) for financial support and scholarships.